\documentstyle[preprint,aps,epsf,tighten]{revtex}
\tightenlines
\begin{document}

\title{TRUE TRANSFORMATIONS OF SPACETIME LENGTHS 
AND\ \ APPARENT\ TRANSFORMATIONS\
OF\ SPATIAL\ AND\ TEMPORAL\ DISTANCES. I. THE THEORY }
\author{Tomislav Ivezi\'{c} \\
%EndAName
\textit{Ru{%
\mbox
 {\it{d}\hspace{-.15em}\rule[1.25ex]{.2em}{.04ex}\hspace{-.05em}}}er Bo\v
{s}kovi\'{c} Institute, P.O.B. 180, 10002 Zagreb, Croatia}\\
\textit{\ ivezic@rudjer.irb.hr}}
\maketitle

It is shown in this paper that the difference between the two forms of
relativity - the ''true transformation (TT) relativity'' and - the
''apparent transformation (AT) relativity'' is essentially caused by the
difference in the concept of \emph{sameness} of a physical system, i.e., of
a physical quantity, for different, relatively moving, observers. In the
''TT relativity'' the same quantity for different inertial frames of
reference is covariantly defined four-dimensional (4D) tensor quantity,
which transforms according to the Lorentz transformation as the TT. In the
''AT relativity'' parts of a 4D tensor quantity are often considered as the
same quantity for different observers, although they correspond to different
quantities in 4D spacetime, and they are not connected by the Lorentz
transformation than by the AT. Then the true transformations of a spacetime
length and the apparent transformations of a spatial distance (the Lorentz
contraction) and of a temporal distance (the usual dilatation of time) are
examined in detail. It is proved that only the true transformations of the
spacetime length are in agreement with the special relativity as the theory
of a 4D spacetime with the pseudo-Euclidean geometry.\medskip

\noindent PACS number(s): 03.30.+p\bigskip

\noindent \textit{Henceforth space by itself, and time by itself, are doomed 
\newline
to fade away into mere shadows and only a kind of union of \newline
the two will preserve an independent reality.} H. Minkowski\medskip

\noindent \textit{A quantity is therefore physically meaningful (in the
sense that it is of the same nature to all observers) if it has tensorial
properties under Lorentz transformations. }F. Rohrlich

\section{INTRODUCTION}

In \cite{ivezic} and \cite{ive2} (see also \cite{ivsc98}) two forms of
relativity are discussed, the ''true transformations (TT) relativity'' and
the ''apparent transformations (AT) relativity.'' The notions of the TT and
the AT are first introduced by Rohrlich \cite{rohrl1}, and, in the same
meaning, but not under that name, discussed in \cite{gamba} too.

\emph{The TT are the transformations of the four-dimensional (4D) spacetime
tensors referring to the same quantity (in 4D spacetime) considered in
different inertial frames of reference (IFRs), or in different
coordinatizations of some IFR.}\textit{\ }An example of the TT are the
covariant Lorentz transformations (LT) of 4D tensor quantities, (see \cite
{Fahn} and \cite{ive2}). Such covariant LT as the TT are the transformations
in 4D spacetime and they transform some 4D tensor quantity $%
Q_{b..}^{a..}(x^{c},x^{d},..)$ from an IFR $S$ to $Q_{b..}^{\prime
a..}(x^{\prime c},x^{\prime d},..)$ in relatively moving IFR $S^{\prime },$
(all parts of the quantity are transformed). Since the ''TT relativity'' is
based on the TT of physical quantities it is obviously the manifestly
covariant formulation of the special relativity. However, there is an
important difference between the usual covariant formulation of the special
relativity and the ''TT relativity.'' Namely, it is considered in the usual
covariant formulation of the special relativity\emph{\ }that the 4D
spacetime tensor quantities, e.g., the electromagnetic field tensor $F^{ab},$
do have well-defined mathematical meaning in the theory but, nevertheless,
the real physical meaning in experiments is assumed to be attributed not to
such 4D spacetime quantities than to some related quantities, e.g., the
electric and magnetic three-vectors (3-vectors) $\mathbf{E}$ and $\mathbf{B,}
$ from the ''3+1'' space and time. In the ''TT relativity'' only one sort of
quantities, the 4D spacetime tensor quantities, do have well-defined meaning
in our 4D spacetime, both, mathematical meaning in the theory, and a real
physical meaning in experiments; \emph{the complete and well-defined
measurement from the ''TT relativity'' viewpoint is such measurement in which%
} \emph{all parts of some 4D quantity are measured. }

In contrast to the TT\emph{\ the AT are not the transformations of 4D
spacetime tensors and they do not refer to the same 4D quantity, but to
different quantities in 4D spacetime. Usually, depending on the used AT,
only a part of a 4D tensor quantity is transformed by the AT, and such a
part of a 4D quantity, when considered in different IFRs (or in different
coordinatizations of some IFR), corresponds to different quantities in 4D
spacetime. Thus, for example, the AT refer to the quantities defined by the
same way of measurement in different IFRs. }An example of\textit{\ }the ''AT
relativity'' is the conventional special relativity based on two Einstein's
postulates and, consequently, on the relativity of simultaneity, on the
synchronous definition of \emph{the spatial length}, i.e., on the AT of the
spatial length (the Lorentz contraction, see \cite
{rohrl1,gamba,ivezic,ive2,ivsc98}), and the AT of \emph{the temporal distance%
} (the conventional dilatation of time), as will be proved in this paper,
and, as shown in \cite{ivezic} (see also \cite{ivsc98}), on the AT of \emph{%
the electric and magnetic 3-vectors} $\mathbf{E}$ and $\mathbf{B}$ (the
conventional transformations of $\mathbf{E}$ and $\mathbf{B,}$ see, e.g., 
\cite{jacks}, Sec.11.10).

In this paper, Sec.2, some general consideration on the two forms of
relativity will be done. In Sec.3 the TT of the spacetime length for - a
moving rod, Sec.3.1, and - a moving clock, Sec.3.2, are exposed. Then in
Sec.4 the AT of - the spatial distance (the Lorentz ''contraction''),
Sec.4.1, and of - the temporal distance (the time ''dilatation''), Sec.4.2,
are considered. Conclusions are given in Sec.5.

\section{GENERAL\ CONSIDERATION\ ON\ THE\ ''TT RELATIVITY'' AND\ THE\ ''AT\
RELATIVITY''}

The main difference between the ''TT relativity'' and the ''AT relativity''
stems from the difference in the concept of \emph{sameness} of a physical
system, i.e., of a physical quantity, for different observers. That concept
actually determines the difference in what is to be understood as a
relativistic theory.

In the ''TT relativity'' as Rohrlich \cite{rohrl1} states: ''The special
theory of relativity is characterized by the group of Lorentz
transformations which describes the way two different observers relate their
observations of \emph{the same physical systems}.'' (my emphasis) Then he
continues with the words taken here as a second motto: ''A quantity is
therefore physically meaningful (in the sense that it is of the same nature
to all observers) if it has tensorial properties under Lorentz
transformations.\textit{'' }Similarly Gamba states in \cite{gamba}\textit{:
''}Special relativity gives us rules to compare results of an experiment
performed by an observer $S$ with results obtained by another observer $%
S^{\prime }$, moving with constant velocity with respect to $S$. It is, of
course, implied that both obsevers are experimenting upon the \emph{same }%
physical system ... ,'' and ''The quantity $A_{\mu }(x_{\lambda },X_{\lambda
})$ for $S$ is the same as the quantity $A_{\mu }^{\prime }(x_{\lambda
}^{\prime },X_{\lambda }^{\prime })$ for $S^{\prime }$ when all the primed
quantities are obtained from the corresponding unprimed quantities through
Lorentz transformations (tensor calculus).''

These examples show that in the the ''TT relativity'' the special relativity
is understood as the theory of 4D spacetime with pseudo-Euclidean geometry.
Quantities of physical interest, both local and nonlocal, are represented by
spacetime tensors, i.e., as covariant quantities, and the laws of physics
are written in a manifestly covariant way as tensorial equations\emph{.}
Such an understanding of the special relativity and of the concept of \emph{%
sameness} of a physical system, i.e., of a physical quantity, is
consistently applied in \cite{ivezic}, \cite{ive2} (see also \cite
{ivsc98,ivescle,ivmom}) by extending the works \cite{rohrl1} and \cite{gamba}
to the relativistic electrodynamics in terms of the introduction of the
four-vectors (4-vectors) $E^{\alpha }$ and $B^{\alpha }$ of the electric and
magnetic field, respectively, and their TT, instead of the usual 3-vectors $%
\mathbf{E}$ and $\mathbf{B}$ and their AT (the conventional transformations
of $\mathbf{E}$ and $\mathbf{B,}$ \cite{jacks}). It has to be noted that
although Rohrlich \cite{rohrl1} and Gamba \cite{gamba} clearly exposed the
concept of sameness of a physical quantity in 4D spacetime they also did not
notice that the usual transformations of the 3-vectors $\mathbf{E}$ and $%
\mathbf{B}$ \cite{jacks} are - the AT, i.e., that $\mathbf{E}$ in $S$ and $%
\mathbf{E}^{\prime }$ in $S^{\prime }$ do not refer to the same 4D tensor
quantity. The covariant formulation of electrodynamics with 4-vectors $%
E^{\alpha }$ and $B^{\alpha }$ is constructed in \cite{ivezic,ive2,ivsc98}
and shown to be equivalent to the usual covariant formulation with the
electromagnetic field tensor $F^{\alpha \beta }.$ Also the covariant
Majorana form of Maxwell's equations is done in \cite{ive2}, while in \cite
{ivezic} (and \cite{ivmom}) the covariant form of the energy-momentum
density tensor $T^{\alpha \beta }$ for the electromagnetic field and the
fully covariant form of a nonlocal quantity, the electromagnetic momentum
4-vector $P_{f}^{\alpha }$, are constructed in terms of the 4-vectors $%
E^{\alpha }$ and $B^{\alpha }.$ It has to be noticed once again that in the
''TT relativity'' only such 4D tensor quantities as are $E^{\alpha }$\ and $%
B^{\alpha },$\ $F^{\alpha \beta },$\ $T^{\alpha \beta },$\ $P_{f}^{\alpha },$%
\ ..., are considered to be well-defined not only mathematically but also
experimentally, as measurable quantities with real physical meaning.

The laws of physics written as tensorial equations with 4D spacetime tensors
in an IFR will have the same form in some other IFR, i.e., in new
coordinates, if new and old coordinates are connected by those coordinate
transformations (the TT) that leave the interval $ds$, and thus the
pseudo-Euclidean geometry of the spacetime, unchanged. (It is explicitly
shown in \cite{ive2} that the AT - the Lorentz contraction, as a coordinate
transformation in 4D spacetime, changes the infinitesimal spacetime distance 
$ds.$) In fact, it is more correct to say that the laws of physics will have
the same form for those coordinate transformations that leave the form,
i.e., the functional dependence, of the metric tensor unchanged\emph{.} Then 
$ds$ will also be unchanged under such coordinate transformations, but,
generally, the reverse does not hold. This means that \emph{in the reference
frames that are connected by such coordinate transformations all physical
phenomena will proceed in the same way, (taking into account the
corresponding initial and boundary conditions), and thus there is no
physical difference between them, what is the content of the principle of
relativity.} \emph{The existence or nonexistence of the group of
transformations that assure the form-invariance of the metric tensor, and
thus also the form-invariance of the covariant equations (physical laws), is
completely determined by the spacetime geometry. Hence, one concludes that
in the ''TT relativity'' the principle of relativity, in contrast to the
Einstein formulation of the special relativity \cite{Einst}, is not a
fundamental principle, than it is a simple consequence of the spacetime
geometry. }

(The geometry of the spacetime is generally defined by the invariant
infinitesimal spacetime distance $ds$ of two neighboring points, 
\begin{equation}
ds^{2}=dx^{a}g_{ab}dx^{b}.  \label{spadist}
\end{equation}
I adopt the following convention with regard to indices. Repeated indices
imply summation. Latin indices $a,b,c,d,...$ are to be read according to the
abstract index notation, see \cite{Wald}, Sec.2.4.. They designate geometric
objects and they run from 0 to 3. Thus $dx^{a,b}$ and $g_{ab},$ and of
course $ds$ (\ref{spadist})$,$ are defined independently of any coordinate
system, e.g., $g_{ab}$ is a second-rank covariant tensor (whose Riemann
curvature tensor $R_{bcd}^{a}$ is everywhere vanishing; the spacetime of
special relativity is a flat spacetime, and this definition includes not
only the IFRs but also the accelerated frames of reference). Greek indices
run from 0 to 3, while latin indices $i,j,k,l,...$ run from 1 to 3, and they
both designate the components of some geometric object in some coordinate
chart, e.g., $x^{\mu }(x^{0},x^{i})$ and $x^{\prime \mu }(x^{\prime
0},x^{\prime i})$ are two coordinate representations of the position
4-vector $x^{a}$ in two different inertial coordinate systems $S$ and $%
S^{\prime },$ and $g_{\mu \nu }$ is the $4\times 4$ matrix of components of $%
g_{ab}$ in some coordinate chart. Let the coordinate transformations from $S$
to $S^{\prime }$ be $x^{\prime \mu }=x^{\prime \mu }(x^{\nu }).$ Then the
metric tensor $g_{\mu \nu }$ transforms according to the law $g_{\mu \nu
}^{\prime }(x^{\prime })=(\partial x^{\alpha }/\partial x^{\prime \mu
})(\partial x^{\beta }/\partial x^{\prime \nu })g_{\mu \nu }(x(x^{\prime
})), $ and if the coordinate transformations are such that they leave the
form, i.e., the functional dependence, of the metric tensor unchanged, then $%
ds$ will necessarily be an invariant quantity under such coordinate
transformations.)

Since the ''TT relativity'' deals on the same footing with all possible
coordinatizations of a chosen reference frame (inertial or accelerated), 
\emph{the second Einstein postulate referred to the constancy of the
coordinate velocity of light also does not hold in the ''TT relativity.''}
Only in Einstein's coordinatization (''e'' coordinatization; when Einstein's
synchronization of distant clocks and cartesian space coordinates $x^{i}$
are used in an IFR S) the coordinate, one-way, speed of light is isotropic
and constant.

Thus \emph{the basic elements of the ''TT relativity,'' as a covariant
formulation of the special relativity, and of the usual Einstein's
formulation of the special relativity, are quite different. Einstein's
formulation is based on two postulates: the principle of relativity and the
constancy of the velocity of light. In the ''TT relativity'' the primary
importance is attributed to the geometry of the spacetime; it is supposed
that the geometry of our 4D spacetime is a pseudo-Euclidean geometry in
which only 4D tensor quantities do have real physical meaning}. (The similar
ideas about the primary importance of the geometry of the spacetime, not
only in the general relativity but also in the special relativity, instead
of Einstein's postulates \cite{Einst}, are expressed in several modern
treatments, e.g., \cite{Logun}.)

Einstein \cite{Einst}, and many others, considered that general laws of
physics must be covariant, but for Einstein, and for the majority of
physicists, such covariance of general laws does not necessarily mean that
the physical quantities of the theory have to be defined in a covariant
manner, as covariant 4D tensor quantities. Thus, for example, Einstein \cite
{Einst} introduced into the special relativity several quantities that are
not covariantly defined and whose transformations are the AT. The examples
are: the synchronously defined spatial length with the AT - the Lorentz
contraction, the temporal distance with the AT - the time dilatation, which
will be discussed here. He also used the 3-vectors $\mathbf{E}$ and $\mathbf{%
B}$ in the formulation of the relativistic electrodynamics$\mathbf{,}$ and
derived their transformations, which are not the LT\ of some well-defined
quantity in 4D spacetime and they do not refer to the same 4D quantity,
i.e., they are also the AT, as shown in \cite{ivezic} (and \cite{ivsc98}). 
\emph{In all cases in which two quantities connected by the AT are
considered to refer to the same physical quantity in 4D spacetime we have},
what Gamba \cite{gamba} calls, \emph{the case of mistaken identity}. To
better explain this issue I quote Gamba's words, \cite{gamba}: ''As far as
relativity is concerned, quantities like $A_{\mu }$ and $\mathcal{A}_{\mu
}^{\prime }$ (they are from different IFRs $S$ and $S^{\prime },$ and \emph{%
they are connected by the AT}, my remark) are different quantities, not
necessarily related to one another. To ask the relation between $\mathcal{A}%
_{\mu }^{\prime }$ and $A_{\mu },$ from the point of view of relativity, is
like asking what is the relation between the measurement of the radius of
the Earth made by an observer $S$ and the measurement of the radius of Venus
made by an observer $S^{\prime }.$ We can certainly take the ratio of the
two measures; what is wrong is the tacit assumption that relativity has
something to do with the problem just because the measurements were made by 
\emph{two} observers.'' (At this point I remark once again that neither
Gamba, despite of such clear understanding of the concept of sameness of a
physical system in 4D spacetime, did not notice that the usual
transformations of the 3-vectors $\mathbf{E}$ and $\mathbf{B}$ are, in fact,
the AT, and in that respect he also dealt with the case of mistaken
identity.)

\section{TRUE TRANSFORMATIONS\ OF THE SPACETIME\ LENGTH}

The whole above consideration is performed in order to emphasize the
importance of the geometry of 4D spacetime in the formulation of special
relativity, and to point out general differences between the ''TT
relativity'' and the ''AT relativity.'' In the following sections these
differences will be illustrated considering some specific examples, the
spacetime length with its TT and then the spatial and temporal distances
with their AT.

\subsection{The spacetime length - for a moving rod}

For the sake of completness we repeat (and in some measure expand) the main
results for the spacetime length, and for the AT of the spatial distance,
that were already found in \cite{ive2}. The invariant spacetime length (the
Lorentz scalar) between two points (events) in 4D spacetime does have
definite physical meaning in the ''TT relativity'' and it is defined as (in
the abstract index notation) 
\begin{equation}
l=(l^{a}g_{ab}l^{b})^{1/2},  \label{length}
\end{equation}
where $l^{a}(l^{b})$ is the distance 4-vector between two events $A$ and $B$%
, $l^{a}=l_{AB}^{a}=x_{B}^{a}-x_{A}^{a}$, $x_{A,B}^{a}$ are the position
4-vectors and $g_{ab}$ is the metric tensor. Using different
coordinatizations of a given reference frame one can find different
expressions, i.e., different representations of the spacetime length $l,$
Eq.(\ref{length}). We shall consider $l$ in two relatively moving IFRs $S$
and $S^{\prime }$ and in two coordinatizations ''e'' and ''r'' in these
IFRs, where ''e'' stands for Einstein's coordinatization in which Einstein's
synchronization of distant clocks and cartesian space coordinates $x^{i}$
are used in an IFR, and where ''r'' stands for ''radio'' coordinatization of
an IFR in which ''everyday'' or ''radio'' synchronization of distant clocks
is used, see \cite{ive2}. (Different synchronizations are determined by the
parameter $\varepsilon $ in the relation $t_{2}=t_{1}+\varepsilon
(t_{3}-t_{1})$, where $t_{1}$ and $t_{3}$ are the times of departure and
arrival, respectively, of the light signal, read by the clock at $A$, and $%
t_{2}$ is the time of reflection at $B$, read by the clock at $B$, that has
to be synchronized with the clock at $A$. In Einstein's synchronization
convention $\varepsilon =1/2$ and the measured coordinate velocity of light
is constant and isotropic. A nice example of a non-standard synchronization
is ''everyday'' or ''radio'' clock synchronization \cite{Leub} in which $%
\varepsilon =0$ and there is an absolute simultaneity; see also \cite{anders}%
).

For further purposes we shall also need a covariant 4D expression for pure
LT when written in geometrical terms, see \cite{Fahn} and \cite{ive2}, 
\begin{equation}
L^{a}{}_{b}\equiv L^{a}{}_{b}(v)=g^{a}{}_{b}-\frac{2u^{a}v_{b}}{c^{2}}+\frac{%
(u^{a}+v^{a})(u_{b}+v_{b})}{c^{2}-u\cdot v},  \label{fah}
\end{equation}
where $u^{a}$ is the proper velocity 4-vector of a frame $S$ with respect to
itself, $u^{a}=cn^{a},$ $n^{a}$ is the unit 4-vector along the $x^{0}$ axis
of the frame $S,$ and $v^{a}$ is the proper velocity 4-vector of $S^{\prime
} $ relative to $S,$ and $u\cdot v=u^{a}v_{a}.$ The form of the covariant 4D
Lorentz transformations $\left( \ref{fah}\right) $ is independent of the
chosen synchronization, i.e., coordinatization of reference frames. With the
use of $\left( \ref{fah}\right) $ the transformation of covariantly defined
physical quantities reduces to the evaluation of invariant scalar products,
e.g., when $L^{a}{}_{b}$ is applied to the position 4-vector $x^{a}$ one
finds (in the abstract index notation) 
\begin{equation}
x^{\prime a}=g^{a}{}_{b}x^{b}+\frac{\left[ n\cdot x-(2\gamma +1)v\cdot
x/c\right] n^{a}+(n\cdot x+v\cdot x/c)v^{a}/c}{1+\gamma },  \label{fahn2}
\end{equation}
where $\gamma =-n\cdot v/c.$

When Einstein's synchronization of distant clocks and cartesian space
coordinates $x^{i}$ are used in an IFR S then, e.g., the geometric object $%
g_{ab}$ is represented by the $4\times 4$ matrix of components of $g_{ab}$
in that coordinate chart, i.e., it is the Minkowski metric tensor $g_{\mu
\nu ,e}=diag(-1,1,1,1).$ Then $L^{a}{}_{b}$ is represented by $L^{\mu
}{}_{\nu ,e},$ the usual expression for pure LT, but with $v_{e}^{i}$ (the
proper velocity 4-vector $v_{e}^{\mu }$ is $v_{e}^{\mu }\equiv dx_{e}^{\mu
}/d\tau =(\gamma _{e}c,\gamma _{e}v_{e}^{i}),$ $d\tau \equiv dt_{e}/\gamma
_{e}$ is the scalar proper-time, and $\gamma _{e}\equiv
(1-v_{e}^{2}/c^{2})^{1/2}$) replacing the components of the ordinary
velocity 3-vector $\mathbf{V.}$ (In the usual form the LT connect two
coordinate representations (in the ''e'' coordinatization) $x_{e}^{\mu
},x_{e}^{\prime \mu }$ of a given event. $x_{e}^{\mu },x_{e}^{\prime \mu }$
refer to two relatively moving IFRs (with the Minkowski metric tensor) $S$
and $S^{\prime },$ 
\begin{eqnarray*}
x_{e}^{\prime \mu } &=&L^{\mu }{}_{\nu ,e}x_{e}^{\nu
},\,L^{0}{}_{0,e}=\gamma _{e},L^{0}{}_{i,e}=L^{i}{}_{0,e}=-\gamma
_{e}v_{e}^{i}/c, \\
L^{i}{}_{j,e} &=&\delta _{j}^{i}+(\gamma _{e}-1)v_{e}^{i}v_{je}/v_{e}^{2}.
\end{eqnarray*}
Since $g_{\mu \nu ,e}$ is a diagonal tensor the space $x_{e}^{i}$ and time $%
t_{e}$ $(x_{e}^{0}\equiv ct_{e})$ components of $x_{e}^{\mu }$ do have their
usual meaning. Then the geometrical quantity $ds^{2}$ (\ref{spadist}) can be
written in terms of its representation $ds_{e}^{2},$ with the separated
spatial and temporal parts, $%
ds^{2}=ds_{e}^{2}=(dx_{e}^{i}dx_{ie})-(dx_{e}^{0})^{2}$, and the same
happens with the spacetime length $l$ (\ref{length})$,$ $%
l^{2}=l_{e}^{2}=(l_{e}^{i}l_{ie})-(l_{e}^{0})^{2}$. Such separation remains
valid in other inertial coordinate systems with the Minkowski metric tensor,
and in $S^{\prime }$ one finds $l^{2}=l_{e}^{\prime 2}=(l_{e}^{\prime
i}l_{ie}^{\prime })-(l_{e}^{\prime 0})^{2},$ where $l_{e}^{\prime \mu }$ in $%
S^{\prime }$ is connected with $l_{e}^{\mu }$ in $S$ by the LT $L^{\mu
}{}_{\nu ,e}$.

Let us also consider the above relations in the ''r'' coordinatization. By
the same construction as in \cite{Leub} we can find the relations between
the base vectors in ''r'' and ''e'' coordinatizations. (We consider, as in 
\cite{Leub} and \cite{ive2} (but now in 4D spacetime), that the spacetime is
endowed with base vectors, the temporal and the spatial base vectors. The
bases $\left\{ e_{\mu }\right\} $, with the base vectors $\left\{
e_{0},e_{i}\right\} $, and $\left\{ r_{\mu }\right\} ,$ with the base
vectors $\left\{ r_{0},r_{i}\right\} ,$ are associated with ''e'' and ''r''
coordinatizations, respectively, of a given IFR.) The connection between the
bases $\left\{ e_{\mu }\right\} $ and $\left\{ r_{\mu }\right\} $ is 
\begin{equation}
r_{0}=e_{0},\;r_{i}=e_{0}+e_{i}.  \label{jedvek}
\end{equation}
Then the metric tensor $g_{ab}$ becomes $g_{\mu \nu ,r}$ with 
\begin{equation}
g_{00,r}=g_{0i,r}=g_{i0,r}=g_{ij,r}(i\neq j)=-1,g_{ii,r}=0.  \label{metener}
\end{equation}
The knowledge of $g_{\mu \nu ,r}$ enable us to find the transformation
matrix between ''r'' and ''e'' coordinatizations.

In \cite{Logun} Logunov derived the expression for the transformation matrix
connecting differentials of physical (how he named it) time and distance $%
dX^{\overline{\mu }}$ with the coordinate ones $dx^{\mu },$ ($dX^{\overline{%
\mu }}=\lambda _{\nu }^{\overline{\mu }}dx^{\nu },$ in his notation, \cite
{Logun} Sec.22.), and by the same matrix $\lambda _{\nu }^{\overline{\mu }}$
he connected a physicaly measurable tensor with the coordinate one. Thus in
the approach of \cite{Logun} there are physical and coordinate quantities
for the same coordinatization of the considered IFR. In my opinion both $dX^{%
\overline{\mu }}$ and $dx^{\nu }$ are equally well ''physical'' and
measurable quantities, and we can interpret that $dX^{\overline{\mu }}$
corresponds to the Einstein coordinatization of a given IFR, while $dx^{\nu
} $ corresponds to some arbitrary coordinatization of the same IFR. Hence,
in my interpretation of Logunov results his matrix $\lambda _{\nu }^{%
\overline{\mu }}$ is, actually, the transformation matrix between some
arbitrary coordinatization and the ''e'' coordinatization. It has to be
noted that although in the Einstein coordinatization the space and time
components of the position 4-vector do have their usual meaning, i.e., as in
the prerelativistic physics, and in $ds_{e}^{2}$ the spatial and temporal
parts are separated, it does not mean that the ''e'' coordinatization does
have some advantage relative to other coordinatizations and that the
quantities in the ''e'' base are more physical.

The elements of $\lambda _{\nu }^{\overline{\mu }}$ \cite{Logun}, which are
different from zero, are $\lambda _{0}^{\overline{0}}=(-g_{00})^{1/2},$ $%
\lambda _{i}^{\overline{0}}=(-g_{0i})(-g_{00})^{-1/2},\quad \lambda _{i}^{%
\overline{i}}=\left[ g_{ii}-(g_{0i})^{2}/g_{00}\right] ^{1/2}.$ We actually
need the inverse transformation $(\lambda _{\nu }^{\overline{\mu }})^{-1}$
(it will be denoted as $T_{\nu }^{\mu }$ to preserve the similarity with the
notation from \cite{ive2}). Then the elements (that are different from zero)
of the matrix $T_{\nu }^{\mu },$ which transforms the ''e'' coordinatization
to the coordinatization determined by the metric tensor $g_{\mu \nu },$ are 
\begin{eqnarray}
T_{0}^{0} &=&(-g_{00})^{-1/2},\quad T_{i}^{0}=(g_{0i})(-g_{00})^{-1}\left[
g_{ii}-(g_{0i})^{2}/g_{00}\right] ^{-1/2},  \nonumber \\
T_{i}^{i} &=&\left[ g_{ii}-(g_{0i})^{2}/g_{00}\right] ^{-1/2}.
\label{lambda}
\end{eqnarray}
Hence $T_{\nu }^{\mu },$ which transforms the ''e'' coordinatization to the
''r'' coordinatization, is found to be $T_{\mu }^{\mu }=-T_{i}^{0}=1,$ and
all other elements of $T_{\nu }^{\mu }$ are $=0.$ Using that $T_{\nu }^{\mu
} $ we find 
\begin{equation}
x_{r}^{\mu }=T_{\nu }^{\mu }x_{e}^{\nu },\quad
x_{r}^{0}=x_{e}^{0}-x_{e}^{1}-x_{e}^{2}-x_{e}^{3},\quad x_{r}^{i}=x_{e}^{i}.
\label{temini}
\end{equation}
The LT $L^{\mu }\,_{\nu ,r}$ in the ''r'' base can be easily found from $(%
\ref{fah})$ and the known $g_{\mu \nu ,r},$ and the elements that are
different from zero are 
\begin{eqnarray}
x_{r}^{\prime \mu } &=&L^{\mu }{}_{\nu ,r}x_{r}^{\nu },\quad
L^{0}{}_{0,r}=K,\quad L^{0}{}_{2,r}=L^{0}{}_{3,r}=K-1,  \nonumber \\
L^{1}{}_{0,r} &=&L^{1}{}_{2,r}=L^{1}{}_{3,r}=(-\beta
_{r}/K),L^{1}{}_{1,r}=1/K,\quad L^{2}{}_{2,r}=L^{3}{}_{3,r}=1,
\end{eqnarray}
where $K=(1+2\beta _{r})^{1/2},$ and $\beta _{r}=dx_{r}^{1}/dx_{r}^{0}$ is
the velocity of the frame $S^{\prime }$ as measured by the frame $S$ (it is
assumed that $S^{\prime }$ is moving relative to $S$ along the common $%
x_{e}^{1},x_{e}^{\prime 1}-$ axes), $\beta _{r}=\beta _{e}/(1-\beta _{e})$
and it ranges as $-1/2\prec \beta _{r}\prec \infty .$ Since $g_{\mu \nu ,r},$
in contrast to $g_{\mu \nu ,e},$ is not a diagonal metric tensor then in $%
ds_{r}^{2}$ the spatial and temporal parts are not separated, and the same
holds for the spacetime length $l,$ see \cite{ive2} for the results in 2D
spacetime. Expressing $dx_{r}^{\mu },$ or $l_{r}^{\mu }$, in terms of $%
dx_{e}^{\mu },$ or $l_{e}^{\mu },$ one finds that $ds_{r}^{2}=ds_{e}^{2},$
and also, $l_{r}^{2}=l_{e}^{2},$ as it must be.

Next we consider the spacetime length in two relatively moving IFRs $S$ and $%
S^{\prime }$ and in two coordinatizations ''e'' and ''r'' in these IFRs,
i.e., we consider it with respect to $\left\{ e_{\mu }\right\} ,\left\{
e_{\mu }^{\prime }\right\} $ and $\left\{ r_{\mu }\right\} ,\left\{ r_{\mu
}^{\prime }\right\} $ bases.

First we consider in short the same example as in \cite{ive2}, i.e., we
consider a particular choice for the 4-vector $l_{AB}^{a}$ (in the usual
''3+1'' picture it corresponds to an object, a rod, that is at rest in an
IFR $S$ and situated along the common $x_{e}^{1},x_{e}^{\prime 1}-$ axes).
For simplicity we work in 2D spacetime and the situation is pictured in
Fig.1.

The base vectors are constructed as in \cite{Leub} and \cite{ive2}, and here
we expose this construction once again for the sake of clearness of the
whole exposition. The temporal base vector $e_{0}$ is the unit vector
directed along the world line of the clock at the origin. The spatial base
vector by definition connects \textit{simultaneous} events, the event
''clock at rest at the origin reads 0 time'' with the event ''clock at rest
at unit distance from the origin reads 0 time,'' and thus it is
synchronization-dependent. The spatial base vector $e_{1}$ connects two
above mentioned simultaneous events when Einstein's synchronization ($%
\varepsilon =1/2$) of distant clocks is used. The temporal base vector $%
r_{0} $ is the same as $e_{0}.$ The spatial base vector $r_{1}$ connects two
above mentioned simultaneous events when ''radio'' clock synchronization ($%
\varepsilon =0$) of distant clocks is used. All the spatial base vectors $%
r_{1},r_{1}^{\prime },..$ are parallel and directed along an
(observer-independent) light line. Hence, two events that are everyday
(''r'') simultaneous in $S$ are also ''r'' simultaneous for all other IFRs.

\begin{figure}
\begin{center}
\epsfxsize = 10cm
\epsfbox{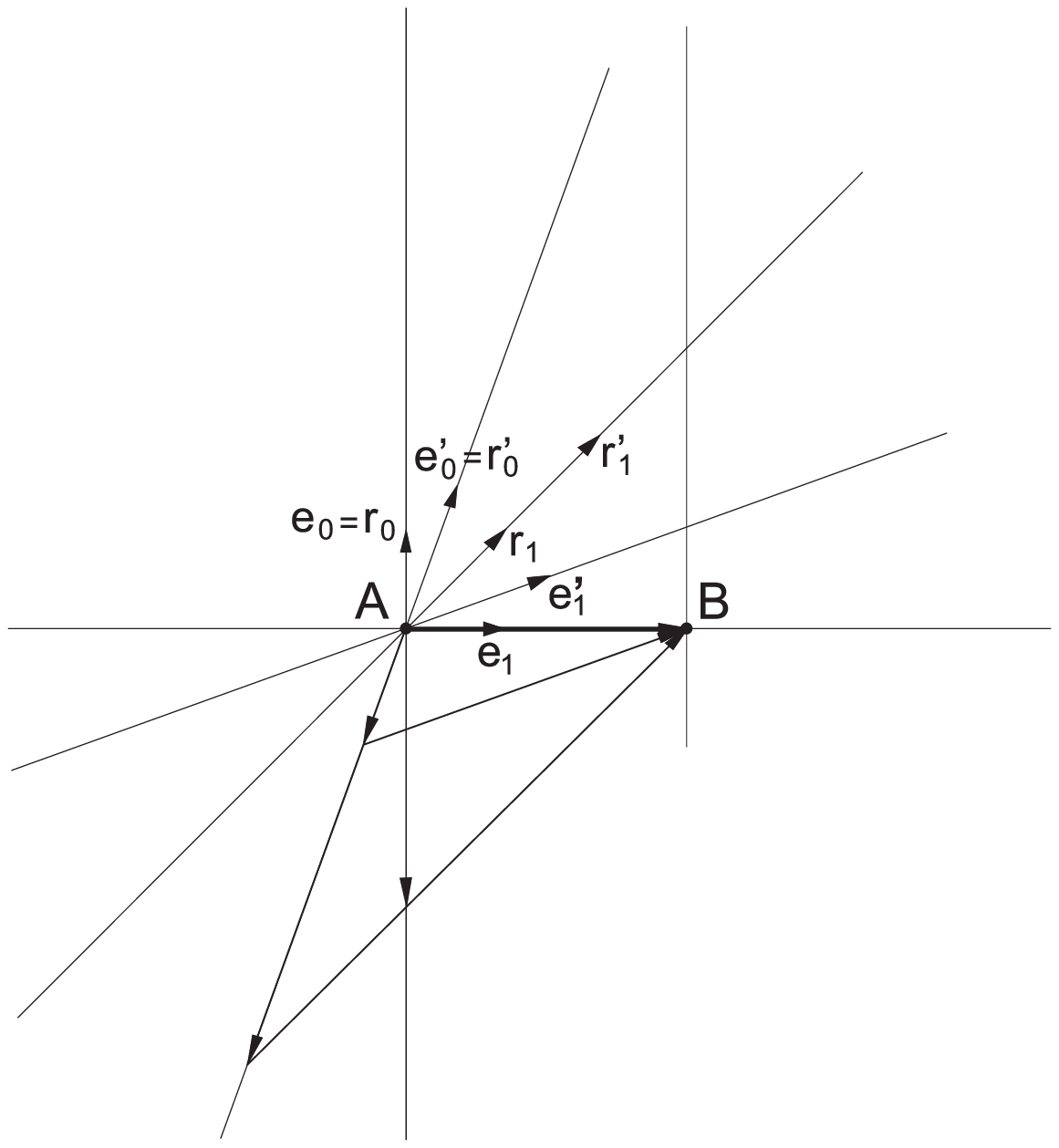}
\caption{ The spacetime length for a moving rod. In the ''TT
relativity'' the same quantity for different observers is the geometrical
quantity, the distance 4-vector $l_{AB}^{a}=x_{B}^{a}-x_{A}^{a};$ only \emph{%
one} quantity in 4D spacetime. It is decomposed with respect to $\left\{
e_{\mu }\right\} ,\left\{ e_{\mu }^{\prime }\right\} $ and $\left\{ r_{\mu
}\right\} ,\left\{ r_{\mu }^{\prime }\right\} $ bases. The bases $\left\{
e_{\mu }\right\} ,\left\{ e_{\mu }^{\prime }\right\} $ refer to Einstein's
coordinatization of two relatively moving IFRs $S$ and $S^{\prime },$ and
the bases $\left\{ r_{\mu }\right\} ,\left\{ r_{\mu }^{\prime }\right\} $
refer to the ''radio'' coordinatization of $S$ and $S^{\prime }.$ $%
l_{AB}^{a} $ corresponds, in the usual ''3+1'' picture, to an object, a rod,
that is at rest in $S$ and situated along the $e_{1}$ base vector. The
representation of $l_{AB}^{a}$ in the $\left\{ e_{\mu }\right\} $ base is $%
l_{AB}^{a}\rightarrow l_{AB,e}^{\mu
}=l_{e}^{0}e_{0}+l_{e}^{1}e_{1}=0e_{0}+l_{0}e_{1},$ in the $\left\{ e_{\mu
}^{\prime }\right\} $ base is $l_{AB}^{a}\rightarrow l_{AB,e}^{\prime \mu
}=-\beta _{e}\gamma _{e}l_{0}e_{0}^{\prime }+\gamma _{e}l_{0}e_{1}^{\prime
}, $ in the $\left\{ r_{\mu }\right\} $ base is $l_{AB}^{a}\rightarrow
l_{AB,r}^{\mu }=-l_{0}r_{0}+l_{0}r_{1},$ and in the $\left\{ r_{\mu
}^{\prime }\right\} $ base is $l_{AB}^{a}\rightarrow l_{AB,r}^{\prime \mu
}=-Kl_{0}r_{0}^{\prime }+(1+\beta _{r})(1/K)l_{0}r_{1}^{\prime },$ where $%
K=(1+2\beta _{r})^{1/2},$ and $\beta _{r}=\beta _{e}/(1-\beta _{e}).\medskip
$}
\end{center}
\end{figure}

The distance 4-vector $l_{AB}^{a}=x_{B}^{a}-x_{A}^{a}$ is decomposed with
respect to $\left\{ e_{\mu }\right\} $ base as 
\begin{equation}
l_{AB}^{a}\rightarrow l_{AB,e}^{\mu
}=l_{e}^{0}e_{0}+l_{e}^{1}e_{1}=0e_{0}+l_{0}e_{1},  \label{elue}
\end{equation}
the temporal part of $l_{AB,e}^{\mu }$ is chosen to be zero. The spacetime
length $l$ is written in the $\left\{ e_{\mu }\right\} $ base as $%
l=l_{e}=(l_{e}^{\mu }l_{\mu e})^{1/2}=(l_{e}^{i}l_{ie})^{1/2}=l_{0},$ as in
the prerelativistic physics; it is in that case a measure of the spatial
distance, i.e., of the rest spatial length of the rod. The observers in all
other IFRs will look at the same events but associating with them different
coordinates and they all obtain the same value $l$ for the spacetime length.
The rest frame of the object, and the simultaneity of the events $A$ and $B$
in it, $l_{e}^{0}=0$, are chosen only to have the connection with the
prerelativistic physics (and the ''AT relativity''), which deals with
''3+1'' quantities and not with 4D quantities. In the ''TT relativity,'' for
the same rod at rest in $S$, we could take another choice for the 4-vector $%
l_{AB}^{a},$ e.g., the choice with $l_{e}^{0}\neq 0.$ The ''TT relativity,''
unlike the nonrelativistic theory and the ''AT relativity,'' is not
interested in the spatial points, the front and the rear ends of the rod,
but \emph{in the events} $A$ and $B$ in the 4D spacetime. The decomposition
of the chosen $l_{AB}^{a}$ relative to the $\left\{ e_{\mu }^{\prime
}\right\} $ base in $S^{\prime },$ (where in the ''3+1'' picture the rod is
moving) is 
\begin{equation}
l_{AB}^{a}\rightarrow l_{AB,e}^{\prime \mu }=-\beta _{e}\gamma
_{e}l_{0}e_{0}^{\prime }+\gamma _{e}l_{0}e_{1}^{\prime }.  \label{eluescr}
\end{equation}
Note that there is a dilatation of the spatial part $l_{e}^{\prime 1}=\gamma
_{e}l_{0}$ with respect to $l_{e}^{1}=l_{0}$ and not the Lorentz contraction
as predicted in the ''AT relativity.'' Hovewer it is clear from the above
discussion that comparison of only spatial parts of the two representations $%
l_{AB,e}^{\mu }$ and $l_{AB,e}^{\prime \mu }$ of the same physical quantity $%
l_{AB}^{a}$ measured in two relatively moving IFRs $S$ and $S^{\prime }$
respectively is physically meaningless in the ''TT relativity.'' The
invariant spacetime length of that object in $S^{\prime }$ is $%
l=l_{e}^{\prime }=l_{0}.$

The distance 4-vector $l_{AB}^{a}=x_{B}^{a}-x_{A}^{a}$ is decomposed with
respect to $\left\{ r_{\mu }\right\} $ base as 
\begin{equation}
l_{AB}^{a}\rightarrow l_{AB,r}^{\mu }=-l_{0}r_{0}+l_{0}r_{1},  \label{disuer}
\end{equation}
and the TT length $l$ is $l=l_{r}=l_{0},$ as it must be. In $S^{\prime }$
and in the $\left\{ r_{\mu }^{\prime }\right\} $ base $l_{AB}^{a}$ is
decomposed as 
\begin{equation}
l_{AB}^{a}\rightarrow l_{AB,r}^{\prime \mu }=-Kl_{0}r_{0}^{\prime }+(1+\beta
_{r})(1/K)l_{0}r_{1}^{\prime }.  \label{disercr}
\end{equation}
If only spatial parts of $l_{AB,r}^{\mu }$ and $l_{AB,r}^{\prime \mu }$ are
compared than one finds that $\infty \succ l_{r}^{\prime 1}\geq l_{0}$ for $%
-1/2\prec \beta _{r}\leq 0$ and $l_{0}\leq l_{r}^{\prime 1}\prec \infty $
for $0\leq \beta _{r}\prec \infty $ , which once again shows that such
comparison is physically meaningless in the ''TT relativity.'' Hovewer the
invariant spacetime length always takes the same value $l=l_{r}^{\prime
}=l_{0}.$ Thus, as also seen from Fig.1, \emph{one and the same geometrical
quantity, the 4-vector }$l_{AB}^{a},$\emph{\ is considered in four different
bases }$\left\{ e_{\mu }\right\} ,\left\{ e_{\mu }^{\prime }\right\}
,\left\{ r_{\mu }\right\} $\emph{\ and }$\left\{ r_{\mu }^{\prime }\right\}
, $\emph{\ where it is represented by its coordinate representations }$%
l_{e}^{\mu },l_{e}^{\prime \mu },l_{r}^{\mu }$\emph{\ and }$l_{r}^{\prime
\mu }$\emph{, respectively.} An important conclusion emerges from the whole
above consideration; \emph{the usual 3D length of a moving object cannot be
defined in the 4D spacetime of the TT relativity in an adequate way, since
it is only the spatial length and not a 4D tensor quantity.}

\subsection{The spacetime length - for a moving clock}

\begin{figure}
\begin{center}
\epsfxsize = 10cm
\epsfbox{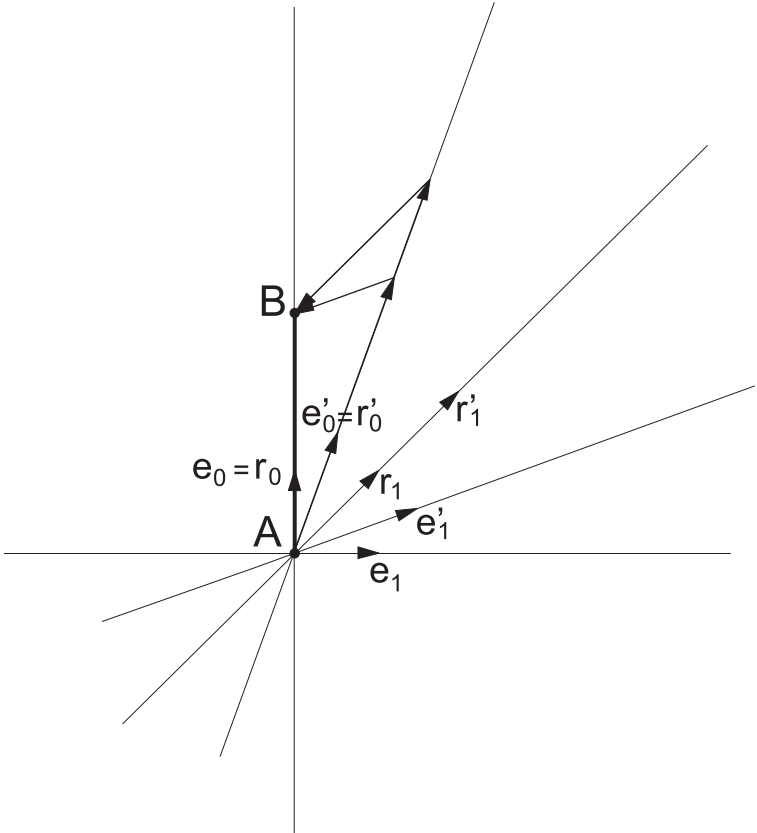}
\caption{The spacetime length for a moving clock. The same
geometrical quantity, the distance 4-vector $l_{AB}^{a}=x_{B}^{a}-x_{A}^{a}$
is decomposed with respect to $\left\{ e_{\mu }\right\} ,\left\{ e_{\mu
}^{\prime }\right\} $ and $\left\{ r_{\mu }\right\} ,\left\{ r_{\mu
}^{\prime }\right\} $ bases. $l_{AB}^{a}$ connects the events $A$ and $B$
(the event $A$ represents the creation of the muon and the event $B$
represents its decay after the lifetime $\tau _{0}$ in $S$)$.$and it is
directed along the $e_{0}$ base vector from the event $A$ toward the event $%
B.$ The representation of $l_{AB}^{a}$ in the $\left\{ e_{\mu }\right\} $
base is $l_{AB}^{a}\rightarrow l_{AB,e}^{\mu }=c\tau _{0}e_{0}+0e_{1},$ in
the $\left\{ e_{\mu }^{\prime }\right\} $ base is $l_{AB}^{a}\rightarrow
l_{AB,e}^{\prime \mu }=\gamma c\tau _{0}e_{0}^{\prime }-\beta \gamma c\tau
_{0}e_{1}^{\prime },$ in the $\left\{ r_{\mu }\right\} $ base is $%
l_{AB}^{a}\rightarrow l_{AB,r}^{\mu }=c\tau _{0}r_{0}+0r_{1},$ and in the $%
\left\{ r_{\mu }^{\prime }\right\} $ base is $l_{AB}^{a}\rightarrow
l_{AB,r}^{\prime \mu }=Kc\tau _{0}r_{0}^{\prime }-\beta K^{-1}c\tau
_{0}r_{1}^{\prime }.\medskip $}
\end{center}
\end{figure}

Another example, i.e., another particular choice for the 4-vector $%
l_{AB}^{a},$ is presented in Fig.2. It clearly reveals the fundamental
difference between the ''TT relativity'' and the ''AT relativity.'' This
example can be interpreted as the well known ''muon experiment,'' but now
considered in the ''TT relativity.'' Again, as in the preceding section, the
spacetime length and $l_{AB}^{a}$ will be examined in two relatively moving
IFRs $S$ and $S^{\prime }$ and in two coordinatizations ''e'' and ''r'' in
these IFRs, i.e., in $\left\{ e_{_{\mu }}\right\} ,\left\{ e_{\mu }^{\prime
}\right\} $ and $\left\{ r_{\mu }\right\} ,\left\{ r_{\mu }^{\prime
}\right\} $ bases. The $S$ frame is chosen to be the rest frame of the muon.
Two events are considered; the event $A$ represents the creation of the muon
and the event $B$ represents its decay after the lifetime $\tau _{0}$ in $S.$
The position 4-vectors of the events $A$ and $B$ in $S,$ which are now taken
to be on the world line of a standard clock that is at rest in the origin of 
$S,$ are decomposed with respect to $\left\{ e_{\mu }\right\} $ base as $%
x_{A}^{a}\rightarrow x_{Ae}^{\mu
}=x_{Ae}^{0}e_{0}+x_{Ae}^{1}e_{1}=0e_{0}+0e_{1},$ and $x_{B}^{a}\rightarrow
x_{Be}^{\mu }=x_{Be}^{0}e_{0}+x_{Be}^{1}e_{1}=c\tau _{0}e_{0}+0e_{1}.$ The
distance 4-vector $l_{AB}^{a}=x_{B}^{a}-x_{A}^{a}$ that connects the events $%
A$ and $B$ is now directed along the $e_{0}$ base vector from the event $A$
toward the event $B.$ It is decomposed in the components in the $\left\{
e_{\mu }\right\} $ base as 
\begin{equation}
l_{AB}^{a}\rightarrow l_{AB,e}^{\mu }=l_{e}^{0}e_{0}+l_{e}^{1}e_{1}=c\tau
_{0}e_{0}+0e_{1}.  \label{timeue}
\end{equation}
We see that in $S$ and the $\left\{ e_{\mu }\right\} $ base the position
4-vectors $x_{A,B}^{a}$ and the distance 4-vector $l_{AB}^{a}$ have only
temporal parts, i.e., $x_{Be}^{1}=x_{Ae}^{1}=l_{e}^{1}=0,$ and the spacetime
length $l$ is $l=l_{e}=(-(l_{e}^{0})^{2})^{1/2}=(-c^{2}\tau _{0}^{2})^{1/2}$%
; it is a measure of the temporal distance in $S$, as in the prerelativistic
physics, and in this case one can speak about the lifetime $\tau _{0}$ of
the muon. In $S^{\prime },$ where this clock (muon) is moving, the position
4-vectors $x_{A}^{a}$ and $x_{B}^{a}$ of the events $A$ and $B,$ the
creation and the decay of the muon respectively, are decomposed with respect
to $\left\{ e_{\mu }^{\prime }\right\} $ base as $x_{A}^{a}\rightarrow
x_{Ae}^{\prime \mu }=x_{Ae}^{\prime 0}e_{0}^{\prime }+x_{Ae}^{\prime
1}e_{1}^{\prime }=0e_{0}^{\prime }+0e_{1}^{\prime },$ and $%
x_{B}^{a}\rightarrow x_{Be}^{\prime \mu }=x_{Be}^{\prime 0}e_{0}^{\prime
}+x_{Be}^{\prime 1}e_{1}^{\prime }=\gamma c\tau _{0}e_{0}^{\prime }-\beta
\gamma c\tau _{0}e_{1}^{\prime },$ and the distance 4-vector $l_{AB}^{a}$ is
decomposed as 
\begin{equation}
l_{AB}^{a}\rightarrow l_{AB,e}^{\prime \mu }=\gamma c\tau _{0}e_{0}^{\prime
}-\beta \gamma c\tau _{0}e_{1}^{\prime }.  \label{timuecrt}
\end{equation}
Now in the $\left\{ e_{\mu }^{\prime }\right\} $ base $l_{AB}^{a}$ contains
not only the temporal part but also the spatial part and again the
comparison of only the temporal parts of the distance 4-vector is physically
meaningless in the ''TT relativity,'' in contrast to the consideration in
the ''AT relativity.'' The notion of the time dilatation, which is in the
''AT relativity'' based on the comparison of only the temporal parts, is
meaningless in the ''TT relativity.'' However the correctly defined quantity
is again the spacetime length, and in $S^{\prime }$ this length is $%
l=l_{e}^{\prime }=l_{e}=(-c^{2}\tau _{0}^{2})^{1/2}.$

In a similar way as above we find that in the ''r'' coordinatization the
position 4-vectors of the events $A$ and $B$ in $S$ are decomposed in the $%
\left\{ r_{\mu }\right\} $ base as $x_{A}^{a}\rightarrow x_{Ar}^{\mu
}=x_{Ar}^{0}r_{0}+x_{Ar}^{1}r_{1}=0r_{0}+0r_{1},$ and $x_{B}^{a}\rightarrow
x_{Br}^{\mu }=c\tau _{0}r_{0}+0r_{1},$ and the distance 4-vector $%
l_{AB}^{a}=x_{B}^{a}-x_{A}^{a}$ is decomposed as 
\begin{equation}
l_{AB}^{a}\rightarrow l_{AB,r}^{\mu }=l_{r}^{0}r_{0}+l_{r}^{1}r_{1}=c\tau
_{0}r_{0}+0r_{1},  \label{timiner}
\end{equation}
and the TT length $l$ is $l=l_{r}=l_{e},$ as it must be. In $S^{\prime }$
and in the $\left\{ r_{\mu }^{\prime }\right\} $ base the position 4-vectors
of the events $A$ and $B$ are $x_{A}^{a}\rightarrow x_{Ar}^{\prime \mu
}=0r_{0}^{\prime }+0r_{1}^{\prime }$ and $x_{B}^{a}\rightarrow
x_{Br}^{\prime \mu }=x_{Br}^{\prime 0}r_{0}^{\prime }+x_{Br}^{\prime
1}r_{1}^{\prime }=(1+2\beta _{r})^{1/2}c\tau _{0}r_{0}^{\prime }-\beta
_{r}(1+2\beta _{r})^{-1/2}c\tau _{0}r_{1}^{\prime }$, and the components $%
l_{r}^{\prime \mu }$ of the distance 4-vector $l_{AB}^{a}$ are equal to the
components $x_{Br}^{\prime \mu }$, i.e., $l_{AB,r}^{\prime \mu
}=x_{Br}^{\prime \mu }.$ Thus 
\begin{equation}
l_{AB}^{a}\rightarrow l_{AB,r}^{\prime \mu }=(1+2\beta _{r})^{1/2}c\tau
_{0}r_{0}^{\prime }-\beta _{r}(1+2\beta _{r})^{-1/2}c\tau _{0}r_{1}^{\prime
}.  \label{timercr}
\end{equation}
Comparing the temporal parts of $l_{AB,r}^{\mu }$ and $l_{AB,r}^{\prime \mu
} $ one finds that $l_{r}^{\prime 0}$ is larger than $l_{r}^{0}=c\tau _{0}$
for $0\prec \beta _{r}\prec \infty ,$ and it is smaller than $l_{r}^{0}$ for 
$-1/2\prec \beta _{r}\prec 0.$ Speaking in the language of the ''AT
relativity'' one could say that for the positive $\beta _{r}$ there is a
''time dilatation'' while for $-1/2\prec \beta _{r}\prec 0$ there is a
''time contraction.'' Since there is no physical reason for the preference
of one coordinatization over the other we once again conclude that the
comparison of only temporal (or spatial) parts of the distance 4-vectors is
physically meaningless. This example nicely reveals the untenability of the
notions from the ''AT relativity'' (the ''Lorentz contraction'' or the
''time dilatation'') in the ''TT relativity'' as the theory of 4D spacetime
with pseudo-Euclidean geometry. However, e.g., the spacetime length is
always correctly defined quantity that takes the same value in both IFRs and
in both coordinatizations, $l=l_{r}^{\prime }=l_{r}=l_{e}^{\prime }=l_{e};$
it can be compared in a physically meaningful sense in the ''TT relativity.''

We see that the geometrical quantities, the 4-vectors, $x_{A,B}^{a},$ $%
l_{AB}^{a}$ have different representations depending on the chosen IFR and
the chosen coordinatization of that IFR, e.g., $x_{e,r}^{\mu
},x_{e,r}^{\prime \mu }$. Although the Einstein coordinatization is
preferred by physicists due to its simplicity and symmetry it is nothing
more ''physical'' than others, e.g., the ''r'' coordinatization.

\section{THE\ AT\ OF\ SPATIAL\ AND\ TEMPORAL\ DISTANCES}

In this section we consider the same two examples as above but now from the
point of view of the conventional, i.e., Einstein's \cite{Einst}
interpretations of \emph{the} \emph{spatial length} of the moving rod and 
\emph{the temporal distance} for the moving clock.

\subsection{The AT of the spatial distance or\quad the Lorentz
''contraction''}

The AT of the spatial distance is already considered in detail in \cite{ive2}
and therefore, here, we only quote the main results and the definitions, and
also illustrate the whole consideration by Fig.3. The same example, a rod at
rest in $S,$ is pictured in Fig.1 when treated in the ''TT relativity,'' and
in Fig.3 when treated in the ''AT relativity.'' It is mentioned in \cite
{ive2} that the synchronous definition of \emph{the spatial length},
introduced by Einstein \cite{Einst}, defines length as \emph{the spatial
distance} between two spatial points on the (moving) object measured by
simultaneity in the rest frame of the observer.

\begin{figure}
\begin{center}
\epsfxsize = 10cm
\epsfbox{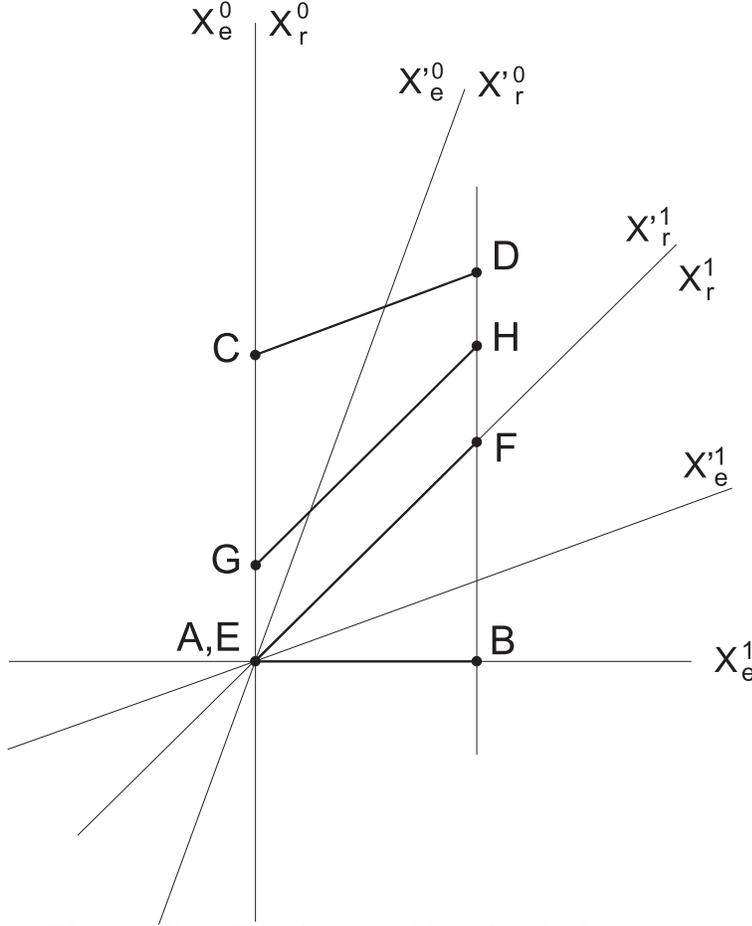}
\caption{The AT of the spatial length - the Lorentz ''contraction''
of the moving rod. The spatial distance $%
l_{ABe}^{1}=x_{Be}^{1}-x_{Ae}^{1}=l_{e}^{1}=l_{0}$ defines in the ''AT
relativity,'' and in the ''e'' base, the spatial length of the rod at rest
in $S,$ while $l_{CDe}^{\prime 1}=x_{De}^{\prime 1}-x_{Ce}^{\prime
1}=l_{e}^{\prime 1}$ is considered in the ''AT relativity,'' and in the
''e'' base, to define the spatial length of the moving rod in $S^{\prime }.$
$l_{e}^{\prime 1}$ and $l_{e}^{1}=l_{0}$ are connected by the formulae for
the Lorentz contraction of the moving rod $l_{e}^{\prime 1}=l_{0}/\gamma
_{e},$ with $t_{Ce}^{\prime }=t_{De}^{\prime }=t_{e}^{\prime }=b$ and $%
t_{Be}=t_{Ae}=t_{e}=a.$ $a$ in $S$ and $b$ in $S^{\prime }$ are not related
by the LT or any other coordinate transformation. Likewise in the ''r''
base, the spatial distance $l_{EFr}^{1}=x_{Fr}^{1}-x_{Er}^{1}$\ defines in
the ''AT relativity'' the spatial length of the rod at rest in $S,$\ while $%
l_{GHr}^{\prime 1}=x_{Hr}^{\prime 1}-x_{Gr}^{\prime 1}$\ defines the spatial
length of the moving rod in $S^{\prime }$. $l_{r}^{\prime 1}=l_{GHr}^{\prime
1}$ and $l_{r}^{1}=l_{EFr}^{1}=l_{0}$ are connected by the formulae for the
Lorentz ''contraction'' of the moving rod in the ''r'' base $l_{r}^{\prime
1}=l_{0}/K$ with $x_{Hr}^{\prime 0}=x_{Gr}^{\prime 0}$ and $%
x_{Fr}^{0}=x_{Er}^{0}.$ In the ''r'' base there is a length dilatation $%
\infty \succ l_{r}^{\prime 1}\succ l_{0}$ for $-1/2\prec \beta _{r}\prec 0$
and the standard ''length contraction'' $l_{0}\succ l_{r}^{\prime 1}\succ 0$
for positive $\beta _{r},$ which clearly shows that the ''Lorentz
contraction'' is not physically correctly defined transformation. In the
''AT relativity'' all four spatial lengths $l_{e}^{1},$ $l_{e}^{\prime 1},$ $%
l_{r}^{1},$ $l_{r}^{\prime 1}$ are considered as the same quantity for
different observers, but, in fact, they are four different quantities in 4D
spacetime, and they are not connected by the Lorentz transformation.\medskip}
\end{center}
\end{figure}

Then, as explained in \cite{ive2}, \emph{the spatial distance }$%
l_{ABe}^{1}=x_{Be}^{1}-x_{Ae}^{1}$\emph{\ defines in the ''AT relativity,''
and in the ''e'' base, the spatial length of the rod at rest in }$S,$ while $%
l_{CDe}^{\prime 1}=x_{De}^{\prime 1}-x_{Ce}^{\prime 1}$ \emph{is considered
in the ''AT relativity,'' and in the ''e'' base, to define the spatial
length of the moving rod in }$S^{\prime },$ see also Fig.3. From these
definitions and considering only the transformation of the spatial part $%
l_{ABe}^{1}$ of the distance 4-vector (in the ''e'' base) $l_{ABe}^{\mu
}=x_{Be}^{\mu }-x_{Ae}^{\mu }=(0,l_{0})$ one finds the relation between $%
l_{e}^{\prime 1}=l_{CDe}^{\prime 1}$ and $l_{e}^{1}=l_{ABe}^{1}=l_{0}$ (see 
\cite{ive2}) as the famous formulae for the Lorentz contraction of the
moving rod 
\begin{equation}
l_{e}^{\prime 1}=x_{De}^{\prime 1}-x_{Ce}^{\prime 1}=l_{0}/\gamma
_{e}=(x_{Be}^{1}-x_{Ae}^{1})(1-\beta _{e}^{2})^{1/2},\quad t_{Ce}^{\prime
}=t_{De}^{\prime },t_{Be}=t_{Ae},  \label{contr}
\end{equation}
where $\beta _{e}=v_{e}/c$, $v_{e}$ is the relative velocity of $S$ and $%
S^{\prime }.$ We see that the spatial lengths $l_{e}^{1}=l_{0}$ and $%
l_{e}^{\prime 1}=l_{0}/\gamma _{e}$ refer not to the same 4D tensor
quantity, as in the ''TT relativity,'' but to two different quantities in 4D
spacetime. These quantities are obtained by the same measurements in $S$ and 
$S^{\prime };$ the spatial ends of the rod are measured simultaneously at
some $t_{e}=a$ in $S$ and also at some $t_{e}^{\prime }=b$ in $S^{\prime },$
and $a$ in $S$ and $b$ in $S^{\prime }$ are not related by the LT or any
other coordinate transformation.

The Lorentz ''contraction'' in the ''r'' coordinatization is also considered
in \cite{ive2} and pictured in Fig.3 here. The spatial ends of the
considered rod, which is at rest in $S,$ must be taken simultaneously in the
''r'' coordinatization too. Thus, in the ''r'' base, they must lie on the
light line, i.e., on the $x_{r}^{1}$ axis (that is along the spatial base
vector $r_{1}$). The simultaneous events $E$ and $F$ (whose spatial parts
correspond to the spatial ends of the rod) are the intersections of the $%
x_{r}^{1}$ axis and the world lines of the spatial ends of the rod. The
events $E$ and $F$ are not the same events as the events $A$ and $B,$
considered in the ''e'' base for the same rod at rest in $S,$ since the
simultaneity of the events is defined in different ways, see Fig.3.
Therefore, in 4D spacetime the spatial length in the ''r'' base $%
l_{r}^{1}=l_{0}$ (with $x_{Fr}^{0}=x_{Er}^{0}$) is not the same 4D quantity
as the spatial length in the ''e'' base $l_{e}^{1}=l_{0}$ (with $%
x_{Be}^{0}=x_{Ae}^{0}).$ Applying the same procedure as in \cite{ive2} one
finds that \emph{in the ''r'' base, the spatial distance }$%
l_{EFr}^{1}=x_{Fr}^{1}-x_{Er}^{1}$\emph{\ defines in the ''AT relativity''
the spatial length of the rod at rest in }$S,$\emph{\ while }$%
l_{GHr}^{\prime 1}=x_{Hr}^{\prime 1}-x_{Gr}^{\prime 1}$\emph{\ defines the
spatial length of the moving rod in }$S^{\prime },$ see Fig.3. Using these
definitions one finds the relation between $l_{r}^{\prime 1}=l_{GHr}^{\prime
1}$ and $l_{r}^{1}=l_{EFr}^{1}=l_{0}$ as the Lorentz ''contraction'' of the
moving rod in the ''r'' base 
\begin{equation}
l_{r}^{\prime 1}=x_{Hr}^{\prime 1}-x_{Gr}^{\prime
1}=l_{0}/K=(1/K)(x_{Fr}^{1}-x_{Er}^{1}),\quad x_{Hr}^{\prime
0}=x_{Gr}^{\prime 0},x_{Fr}^{0}=x_{Er}^{0}.  \label{det}
\end{equation}
In contrast to the ''e'' coordinatization we find that in the ''r'' base
there is a length dilatation $\infty \succ l_{r}^{\prime 1}\succ l_{0}$ for $%
-1/2\prec \beta _{r}\prec 0$ and the standard length ''contraction'' $%
l_{0}\succ l_{r}^{\prime 1}\succ 0$ for positive $\beta _{r},$ which clearly
shows that the ''Lorentz contraction'' is not physically correctly defined
transformation.

At the beginning of Sec.2 we have stated that the main difference between
the ''TT relativity'' and the ''AT relativity'' stems from the difference in
the concept of \emph{sameness} of a physical quantity for different
observers. This statement becomes clear when we compare Fig.1 and Fig. 3 and
the considerations performed in Sec.3.1. and in this one. From Fig.1 and
Sec.3.1. we see that \emph{in the ''TT relativity'' the same quantity for
different observers is the geometrical quantity, the 4-vector }$l_{AB}^{a};$%
\emph{\ only one quantity in 4D spacetime.} On the other hand from Fig.3 and
the discussion in this section we see that in the ''AT relativity'' all four
spatial lengths $l_{e}^{1},$ $l_{e}^{\prime 1},$ $l_{r}^{1},$ $l_{r}^{\prime
1}$ are considered as the same quantity for different observers, but they
are actually four different quantities in 4D spacetime.

Thus we conclude that - \emph{the Lorentz contraction is the transformation
that connects different quantities (in 4D spacetime) in different IFRs,
which shows that it belongs to - the AT. }

\subsection{The AT of the temporal distance or the time ''dilatation''}

In the ''AT relativity'' one can speak not only about the spatial distance
but also about the temporal distance, and they are considered as well
defined quantities. In Fig.4 we present the same ''muon experiment'' as in
Fig.2.

\begin{figure}
\begin{center}
\epsfxsize = 10cm
\epsfbox{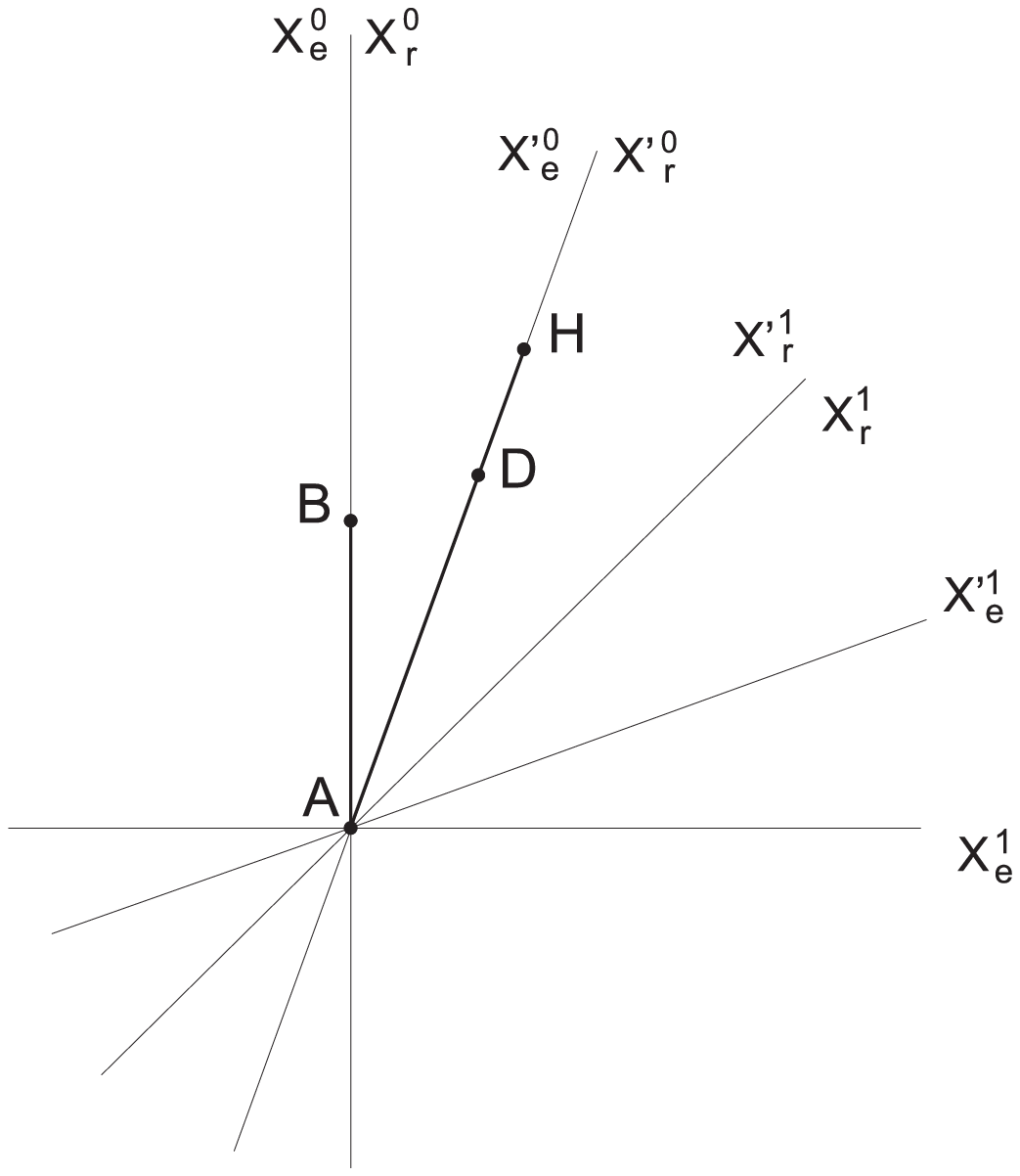}
\caption{The AT of the temporal distance - the ''dilatation'' of    
time for the moving clock. The temporal distance $l_{ABe}^{0}=l_{e}^{0}$ 
defines in the ''AT relativity,'' and in the ''e'' base, the muon lifetime
at rest, while $l_{ADe}^{\prime 0}=l_{e}^{\prime 0}$ is considered in the
''AT relativity,'' and in the ''e'' base, to define the lifetime of the    
moving muon in $S^{\prime }.$ The quantities $l_{e}^{\prime 0}$ and $%       
l_{e}^{0}$ are connected by the relation for the time dilatation, $%        
l_{e}^{\prime 0}/c=t_{e}^{\prime }=\gamma _{e}l_{e}^{0}/c=\tau _{0}(1-\beta
_{e}^{2})^{-1/2},$ with $x_{Be}^{1}=x_{Ae}^{1}.$ Likewise, the temporal    
distance $l_{ABr}^{0}=l_{r}^{0}$ defines in the ''AT relativity,'' and in
the ''r'' base, the muon lifetime at rest, while $l_{AHr}^{\prime
0}=l_{r}^{\prime 0}$ is considered in the ''AT relativity,'' and in the  
''r'' base, to define the lifetime of the moving muon in $S^{\prime }.$ $% 
l_{r}^{\prime 0}$ and $l_{r}^{0}$ are connected by the relation for the time
''dilatation'' in the ''r'' base $l_{r}^{\prime 0}=Kl_{r}^{0}=(1+2\beta
_{r})^{1/2}c\tau _{0}.$ The temporal separation $l_{r}^{\prime 0}$ in $%
S^{\prime },$ where the clock is moving, is smaller - ''time contraction'' -
than the temporal separation $l_{r}^{0}=c\tau _{0}$ in $S,$ where the clock
is at rest, for $-1/2\prec \beta _{r}\prec 0,$ and it is larger - ''time    
dilatation'' - for $0\prec \beta _{r}\prec \infty $. The ''AT relativity''
considers the temporal distances $l_{e}^{0},$\emph{\ }$l_{e}^{\prime 0},$% 
\emph{\ }$l_{r}^{0},$\emph{\ }$l_{r}^{\prime 0}$ as the same quantity for     
different observers. However these temporal distances are really different
quantities in 4D spacetime, and they are not connected by the Lorentz         
transformations.}
\end{center}
\end{figure}

Instead of to work with geometrical quantities $x_{A,B}^{a},l_{AB}^{a}$ and $%
l$ one deals, in the ''AT relativity,'' only with the spatial, or temporal,
parts of their coordinate representations $x_{Ae,r}^{\mu },x_{Be,r}^{\mu }$
and $l_{e,r}^{\mu }.$ First the ''e'' coordinatization, which is almost
always used in the ''AT relativity,'' is considered here. In 4D (at us 2D)
spacetime and in the ''e'' coordinatization the events $A$ and $B$ are again
on the world line of a muon that is at rest in $S$. The distance 4-vector
(in the ''e'' base) is $l_{ABe}^{\mu }=x_{Be}^{\mu }-x_{Ae}^{\mu }=(c\tau
_{0},0).$ Further one uses the Lorentz transformations to express $%
x_{Ae}^{\prime \mu },$ $x_{Be}^{\prime \mu },$ and $l_{ABe}^{\prime \mu }$
in $S^{\prime }$, in which the muon is moving, in terms of the corresponding
quantities in $S$. This procedure yields $x_{A,Be}^{\prime
0}=ct_{A,Be}^{\prime }=\gamma _{e}(ct_{A,Be}-\beta _{e}x_{A,Be}^{1}),$ and $%
x_{A,Be}^{\prime 1}=\gamma _{e}(\beta _{e}ct_{A,Be}-x_{A,Be}^{1}),$ whence 
\begin{eqnarray*}
l_{ABe}^{\prime 0} &=&ct_{Be}^{\prime }-ct_{Ae}^{\prime }=\gamma
_{e}(ct_{Be}-ct_{Ae})-\gamma _{e}\beta _{e}(x_{Be}^{1}-x_{Ae}^{1}) \\
l_{ABe}^{\prime 1} &=&x_{Be}^{\prime 1}-x_{Ae}^{\prime 1}=\gamma
_{e}(x_{Be}^{1}-x_{Ae}^{1})-\gamma _{e}\beta _{e}(ct_{Be}-ct_{Ae}).
\end{eqnarray*}
Now comes the main difference between the two forms of relativity. Instead
of to work with 4D tensor quantities and their LT (as in the ''TT
relativity'') in the ''AT relativity'' the temporal part alone $%
l_{ABe}^{0}=c\tau _{0}$ of $l_{ABe}^{\mu }$ is considered as a well-defined
quantity, and it defines the muon lifetime at rest. The existence of the
spatial part $l_{ABe}^{1}$ of $l_{ABe}^{\mu }$ is in this case completely
neglected; one forgets the transformation of the spatial part $l_{ABe}^{1}$
and considers only the transformation of the temporal part $l_{ABe}^{0}.$
However, in the 4D (at us 2D) spacetime such an assumption means that in $%
S^{\prime }$\ one actually does not consider the same events $A$\ and $B$\
as in $S$\ but some other two events $C$\ and $D$ (in fact, in this specific
example, the events $A$\ and\emph{\ }$D$), see Fig.4. Then in the above
transformation for $l_{ABe}^{0}$ one has to write $x_{De}^{\prime
0}-x_{Ae}^{\prime 0}=l_{ADe}^{\prime 0}$ instead of $x_{Be}^{\prime
0}-x_{Ae}^{\prime 0}=l_{ABe}^{\prime 0}.$ As we said the temporal distance $%
l_{ABe}^{0}=l_{e}^{0}$ defines in the ''AT relativity,'' and in the ''e''
base, the muon lifetime at rest, while $l_{ADe}^{\prime 0}=l_{e}^{\prime 0}$
is considered in the ''AT relativity,'' and in the ''e'' base, to define the
lifetime of the moving muon in $S^{\prime }.$ Taking that $%
x_{Be}^{1}=x_{Ae}^{1}$ in the equation for $l_{e}^{\prime 0}$ one finds the
well-known relation for the time dilatation, 
\begin{equation}
l_{e}^{\prime 0}/c=t_{e}^{\prime }=\gamma _{e}l_{e}^{0}/c=\tau _{0}(1-\beta
_{e}^{2})^{-1/2},\text{ with }x_{Be}^{1}=x_{Ae}^{1}.  \label{tidil}
\end{equation}
As seen from Fig.4 the relation (\ref{tidil}) connects two different
quantities (in 4D spacetime) - the temporal parts of $l_{ABe}^{\mu }$ and $%
l_{ABe}^{\prime \mu },$ i.e., the temporal parts of the $\left\{ e_{\mu
}\right\} $ and $\left\{ e_{\mu }^{\prime }\right\} $ representations of the
distance 4-vector $l_{AB}^{a}$; the quantities $l_{e}^{0}$\emph{\ and }$%
l_{e}^{\prime 0}$\emph{\ are different and, in fact, independent quantities
in 4D spacetime, which are not connected by the Lorentz transformation.}

In the ''r'' coordinatization, see Fig.4., the $\left\{ r_{\mu }\right\} $
representation of the distance 4-vector is $l_{ABr}^{\mu }=x_{Br}^{\mu
}-x_{Ar}^{\mu }=(c\tau _{0},0).$ By means of the LT (\ref{fah}), when
written in the $\left\{ r_{\mu }\right\} $ base, we transform $l_{ABr}^{\mu
} $ to $l_{ABr}^{\prime \mu }$ and find the relations 
\begin{eqnarray*}
l_{ABr}^{\prime 0} &=&x_{Br}^{\prime 0}-x_{Ar}^{\prime
0}=K(x_{Br}^{0}-x_{Ar}^{0})=Kc\tau _{0}, \\
l_{ABr}^{\prime 1} &=&x_{Br}^{\prime 1}-x_{Ar}^{\prime 1}=(-\beta
_{r}/K)(x_{Br}^{0}-x_{Ar}^{0})+(1/K)(x_{Br}^{1}-x_{Ar}^{1}).
\end{eqnarray*}
Further, in the ''r'' base, one again forgets the transformation of the
spatial part $l_{ABr}^{1}$ of $l_{ABr}^{\mu }.$ In the same way as in the
''e'' base, in 4D (at us 2D) spacetime, such an assumption means that in $%
S^{\prime }$ one actually does not consider the same events $A$ and $B$ as
in $S$ but some other two events $G$ and $H,$ (in this particular example,
the events $A$\ and\emph{\ }$H$), see Fig.4. Then in the above
transformation for $l_{ABr}^{0}=l_{r}^{0}$ one has to write $l_{AHr}^{\prime
0}$ instead of $l_{ABr}^{\prime 0};$ the temporal distance $%
l_{ABr}^{0}=l_{r}^{0}$ defines in the ''AT relativity,'' and in the ''r''
base, the muon lifetime at rest, while $l_{AHr}^{\prime 0}=l_{r}^{\prime 0}$
is considered in the ''AT relativity,'' and in the ''r'' base, to define the
lifetime of the moving muon in $S^{\prime }.$ The relation for the time
''dilatation'' in the ''r'' base becomes 
\begin{equation}
l_{r}^{\prime 0}=Kl_{r}^{0}=(1+2\beta _{r})^{1/2}c\tau _{0}.  \label{tider}
\end{equation}
Again, as seen from Fig.4, the relation (\ref{tider}) connects two different
quantities (in 4D spacetime) - the temporal parts of $l_{ABr}^{\mu }$ and $%
l_{ABr}^{\prime \mu },$ i.e., the temporal parts of the $\left\{ r_{\mu
}\right\} $ and $\left\{ r_{\mu }^{\prime }\right\} $ representations of the
distance 4-vector $l_{AB}^{a}$; the quantities $l_{r}^{0}$\emph{\ and }$%
l_{r}^{\prime 0}$\emph{\ are different and, in fact, independent quantities
in 4D spacetime, which are not connected by the Lorentz transformation.} We
see from (\ref{tider}) that the temporal separation $l_{r}^{\prime 0}$ in $%
S^{\prime },$ where the clock is moving, is smaller - ''time contraction'' -
than the temporal separation $l_{r}^{0}=c\tau _{0}$ in $S,$ where the clock
is at rest, for $-1/2\prec \beta _{r}\prec 0,$ and it is larger - ''time
dilatation'' - for $0\prec \beta _{r}\prec \infty $. In numerous papers and
books the time dilatation given by (\ref{tidil}) is considered as a
fundamental relativistic effect, but as shown by the relation (\ref{tider})
the effect depends on the chosen coordinatizations, and as such cannot be
the fundamental effect. In addition, we point out that from the point of
view of the ''TT relativity'' such transformations that transform only some
parts of 4D tensor quantities, and completely neglect the transformations of
the remaining parts, are physically meaningless.

From Fig.2 and Sec.3.2. we see that \emph{in the ''TT relativity'' the same
quantity for different observers is the geometrical quantity, the 4-vector }$%
l_{AB}^{a};$\emph{\ only one quantity in 4D spacetime.} However from Fig.4
and this section we reveal that \emph{in the ''AT relativity'' different
quantities in 4D spacetime, the temporal distances }$l_{e}^{0},$\emph{\ }$%
l_{e}^{\prime 0},$\emph{\ }$l_{r}^{0},$\emph{\ }$l_{r}^{\prime 0},$\emph{\
are considered as the same quantity for different observers.}

One concludes from the above discussion that \emph{both the Lorentz
''contraction'' and the time ''dilatation'' are the transformations
connecting different quantities (in 4D spacetime) in different IFRs, and
they both change the infinitesimal spacetime distance }$ds,$\emph{\ i.e.,
the pseudo-Euclidean geometry of the 4D spacetime }(this is explicitly shown
in \cite{ive2} for the Lorentz ''contraction,'' and also it can be easily
shown for the time ''dilatation''). \emph{Such characteristics of the
Lorentz ''contraction'' and the time ''dilatation'' as the coordinate
transformations show that both transformations belong to - the AT.}\textit{\ 
}

\section{CONCLUSIONS\ AND\ DISCUSSION}

As shown in this paper the two forms of relativity the ''TT relativity'' and
the ''AT relativity'' are substantially different theories. The concept of 
\emph{sameness} of a physical quantity for different observers makes clear
distinction between the two considered forms of relativity. In the ''TT
relativity'' the same quantity for different observers is the geometrical
quantity, only one quantity in 4D spacetime; in the case considered in this
paper it is the 4-vector $l_{AB}^{a},$\ (or the spacetime length $l$, the
Lorentz scalar), as seen from Figs.1 and 2. In contrast to the ''TT
relativity'' the traditionally used ''AT relativity,'' considers different
spatial lengths $l_{e}^{1},$ $l_{e}^{\prime 1},$ $l_{r}^{1},$ $l_{r}^{\prime
1},$ Fig.3 (the temporal distances $l_{e}^{0},$\emph{\ }$l_{e}^{\prime 0},$%
\emph{\ }$l_{r}^{0},$\emph{\ }$l_{r}^{\prime 0},$ Fig.4) as the same
quantity for different observers. However, as seen from Fig.3 (Fig.4) the
spatial lengths $l_{e}^{1},$ $l_{e}^{\prime 1},$ $l_{r}^{1},$ $l_{r}^{\prime
1}$ (the temporal distances $l_{e}^{0},$\emph{\ }$l_{e}^{\prime 0},$\emph{\ }%
$l_{r}^{0},$\emph{\ }$l_{r}^{\prime 0}$) are really different quantities in
4D spacetime, which are not connected by the Lorentz transformations $%
L^{a}{}_{b}$ (\ref{fah})$,$ i.e., the quantities $l_{e}^{1}$ and $%
l_{e}^{\prime 1}$ ($l_{e}^{0}$ and $l_{e}^{\prime 0}$) are not connected by $%
L^{\mu }\,_{\nu ,e},$ nor $l_{r}^{1}$ and $l_{r}^{\prime 1}$ ($l_{r}^{0}$
and $l_{r}^{\prime 0}$) are connected by $L^{\mu }\,_{\nu ,r}$ ($L^{\mu
}\,_{\nu ,e}$ and $L^{\mu }\,_{\nu ,r}$ are the representations of $%
L^{a}{}_{b}$ in the ''e'' and ''r'' coordinatizations, see Sec.3). Also,
neither the quantities $l_{e}^{1}$ and $l_{r}^{1},$ etc., are connected by
the coordinate transformation $T_{\nu }^{\mu },$ which transforms the ''e''
coordinatization to the ''r'' coordinatization, see Sec.3.

The consideration of the spacetime length (the ''TT relativity'') Sec.3, and
of the spatial and temporal distances as well-defined quantities (the ''AT
relativity'') Sec.4, reveals that only the ''TT relativity'' is in a
complete agreement with the special relativity as the theory of 4D spacetime
with pseudo-Euclidean geometry; when the 4D structure of our spacetime is
correctly taken into account as in the ''TT relativity'' then there is no
place either for the Lorentz ''contraction'' or for the time ''dilatation,''
i.e., there is no place for the ''AT relativity.''\pagebreak

\noindent FIGURE CAPTIONS\medskip

\noindent Fig.1. The spacetime length for a moving rod. In the ''TT
relativity'' the same quantity for different observers is the geometrical
quantity, the distance 4-vector $l_{AB}^{a}=x_{B}^{a}-x_{A}^{a};$ only \emph{%
one} quantity in 4D spacetime. It is decomposed with respect to $\left\{
e_{\mu }\right\} ,\left\{ e_{\mu }^{\prime }\right\} $ and $\left\{ r_{\mu
}\right\} ,\left\{ r_{\mu }^{\prime }\right\} $ bases. The bases $\left\{
e_{\mu }\right\} ,\left\{ e_{\mu }^{\prime }\right\} $ refer to Einstein's
coordinatization of two relatively moving IFRs $S$ and $S^{\prime },$ and
the bases $\left\{ r_{\mu }\right\} ,\left\{ r_{\mu }^{\prime }\right\} $
refer to the ''radio'' coordinatization of $S$ and $S^{\prime }.$ $%
l_{AB}^{a} $ corresponds, in the usual ''3+1'' picture, to an object, a rod,
that is at rest in $S$ and situated along the $e_{1}$ base vector. The
representation of $l_{AB}^{a}$ in the $\left\{ e_{\mu }\right\} $ base is $%
l_{AB}^{a}\rightarrow l_{AB,e}^{\mu
}=l_{e}^{0}e_{0}+l_{e}^{1}e_{1}=0e_{0}+l_{0}e_{1},$ in the $\left\{ e_{\mu
}^{\prime }\right\} $ base is $l_{AB}^{a}\rightarrow l_{AB,e}^{\prime \mu
}=-\beta _{e}\gamma _{e}l_{0}e_{0}^{\prime }+\gamma _{e}l_{0}e_{1}^{\prime
}, $ in the $\left\{ r_{\mu }\right\} $ base is $l_{AB}^{a}\rightarrow
l_{AB,r}^{\mu }=-l_{0}r_{0}+l_{0}r_{1},$ and in the $\left\{ r_{\mu
}^{\prime }\right\} $ base is $l_{AB}^{a}\rightarrow l_{AB,r}^{\prime \mu
}=-Kl_{0}r_{0}^{\prime }+(1+\beta _{r})(1/K)l_{0}r_{1}^{\prime },$ where $%
K=(1+2\beta _{r})^{1/2},$ and $\beta _{r}=\beta _{e}/(1-\beta _{e}).\medskip 
$

\noindent Fig.2. The spacetime length for a moving clock. The same
geometrical quantity, the distance 4-vector $l_{AB}^{a}=x_{B}^{a}-x_{A}^{a}$
is decomposed with respect to $\left\{ e_{\mu }\right\} ,\left\{ e_{\mu
}^{\prime }\right\} $ and $\left\{ r_{\mu }\right\} ,\left\{ r_{\mu
}^{\prime }\right\} $ bases. $l_{AB}^{a}$ connects the events $A$ and $B$
(the event $A$ represents the creation of the muon and the event $B$
represents its decay after the lifetime $\tau _{0}$ in $S$)$.$and it is
directed along the $e_{0}$ base vector from the event $A$ toward the event $%
B.$ The representation of $l_{AB}^{a}$ in the $\left\{ e_{\mu }\right\} $
base is $l_{AB}^{a}\rightarrow l_{AB,e}^{\mu }=c\tau _{0}e_{0}+0e_{1},$ in
the $\left\{ e_{\mu }^{\prime }\right\} $ base is $l_{AB}^{a}\rightarrow
l_{AB,e}^{\prime \mu }=\gamma c\tau _{0}e_{0}^{\prime }-\beta \gamma c\tau
_{0}e_{1}^{\prime },$ in the $\left\{ r_{\mu }\right\} $ base is $%
l_{AB}^{a}\rightarrow l_{AB,r}^{\mu }=c\tau _{0}r_{0}+0r_{1},$ and in the $%
\left\{ r_{\mu }^{\prime }\right\} $ base is $l_{AB}^{a}\rightarrow
l_{AB,r}^{\prime \mu }=Kc\tau _{0}r_{0}^{\prime }-\beta K^{-1}c\tau
_{0}r_{1}^{\prime }.\medskip $

\noindent Fig.3. The AT of the spatial length - the Lorentz ''contraction''
of the moving rod. The spatial distance $%
l_{ABe}^{1}=x_{Be}^{1}-x_{Ae}^{1}=l_{e}^{1}=l_{0}$ defines in the ''AT
relativity,'' and in the ''e'' base, the spatial length of the rod at rest
in $S,$ while $l_{CDe}^{\prime 1}=x_{De}^{\prime 1}-x_{Ce}^{\prime
1}=l_{e}^{\prime 1}$ is considered in the ''AT relativity,'' and in the
''e'' base, to define the spatial length of the moving rod in $S^{\prime }.$ 
$l_{e}^{\prime 1}$ and $l_{e}^{1}=l_{0}$ are connected by the formulae for
the Lorentz contraction of the moving rod $l_{e}^{\prime 1}=l_{0}/\gamma
_{e},$ with $t_{Ce}^{\prime }=t_{De}^{\prime }=t_{e}^{\prime }=b$ and $%
t_{Be}=t_{Ae}=t_{e}=a.$ $a$ in $S$ and $b$ in $S^{\prime }$ are not related
by the LT or any other coordinate transformation. Likewise in the ''r''
base, the spatial distance $l_{EFr}^{1}=x_{Fr}^{1}-x_{Er}^{1}$\ defines in
the ''AT relativity'' the spatial length of the rod at rest in $S,$\ while $%
l_{GHr}^{\prime 1}=x_{Hr}^{\prime 1}-x_{Gr}^{\prime 1}$\ defines the spatial
length of the moving rod in $S^{\prime }$. $l_{r}^{\prime 1}=l_{GHr}^{\prime
1}$ and $l_{r}^{1}=l_{EFr}^{1}=l_{0}$ are connected by the formulae for the
Lorentz ''contraction'' of the moving rod in the ''r'' base $l_{r}^{\prime
1}=l_{0}/K$ with $x_{Hr}^{\prime 0}=x_{Gr}^{\prime 0}$ and $%
x_{Fr}^{0}=x_{Er}^{0}.$ In the ''r'' base there is a length dilatation $%
\infty \succ l_{r}^{\prime 1}\succ l_{0}$ for $-1/2\prec \beta _{r}\prec 0$
and the standard ''length contraction'' $l_{0}\succ l_{r}^{\prime 1}\succ 0$
for positive $\beta _{r},$ which clearly shows that the ''Lorentz
contraction'' is not physically correctly defined transformation. In the
''AT relativity'' all four spatial lengths $l_{e}^{1},$ $l_{e}^{\prime 1},$ $%
l_{r}^{1},$ $l_{r}^{\prime 1}$ are considered as the same quantity for
different observers, but, in fact, they are four different quantities in 4D
spacetime, and they are not connected by the Lorentz transformation.\medskip

\noindent Fig.4. The AT of the temporal distance - the ''dilatation'' of
time for the moving clock. The temporal distance $l_{ABe}^{0}=l_{e}^{0}$
defines in the ''AT relativity,'' and in the ''e'' base, the muon lifetime
at rest, while $l_{ADe}^{\prime 0}=l_{e}^{\prime 0}$ is considered in the
''AT relativity,'' and in the ''e'' base, to define the lifetime of the
moving muon in $S^{\prime }.$ The quantities $l_{e}^{\prime 0}$ and $%
l_{e}^{0}$ are connected by the relation for the time dilatation, $%
l_{e}^{\prime 0}/c=t_{e}^{\prime }=\gamma _{e}l_{e}^{0}/c=\tau _{0}(1-\beta
_{e}^{2})^{-1/2},$ with $x_{Be}^{1}=x_{Ae}^{1}.$ Likewise, the temporal
distance $l_{ABr}^{0}=l_{r}^{0}$ defines in the ''AT relativity,'' and in
the ''r'' base, the muon lifetime at rest, while $l_{AHr}^{\prime
0}=l_{r}^{\prime 0}$ is considered in the ''AT relativity,'' and in the
''r'' base, to define the lifetime of the moving muon in $S^{\prime }.$ $%
l_{r}^{\prime 0}$ and $l_{r}^{0}$ are connected by the relation for the time
''dilatation'' in the ''r'' base $l_{r}^{\prime 0}=Kl_{r}^{0}=(1+2\beta
_{r})^{1/2}c\tau _{0}.$ The temporal separation $l_{r}^{\prime 0}$ in $%
S^{\prime },$ where the clock is moving, is smaller - ''time contraction'' -
than the temporal separation $l_{r}^{0}=c\tau _{0}$ in $S,$ where the clock
is at rest, for $-1/2\prec \beta _{r}\prec 0,$ and it is larger - ''time
dilatation'' - for $0\prec \beta _{r}\prec \infty $. The ''AT relativity''
considers the temporal distances $l_{e}^{0},$\emph{\ }$l_{e}^{\prime 0},$%
\emph{\ }$l_{r}^{0},$\emph{\ }$l_{r}^{\prime 0}$ as the same quantity for
different observers. However these temporal distances are really different
quantities in 4D spacetime, and they are not connected by the Lorentz
transformations.

\end{document}